\documentclass[twocolumn,english,aps,pra,showpacs,superscriptaddress,amsmath,amsfonts]{revtex4-1}

\usepackage{epsfig}
\psfigdriver{dvips}
\usepackage{graphicx}
\usepackage{color}
\usepackage{gensymb}
\newcommand{\beq}{\begin{equation}}
\newcommand{\eeq}{\end{equation}}
\newcommand{\bea}{\begin{eqnarray}}
\newcommand{\eea}{\end{eqnarray}}

\usepackage{bm, bbm}      
\usepackage{curves}
\usepackage{epic}
\usepackage{wasysym}
\usepackage{subfigure}
\usepackage{comment}

\usepackage{amssymb}

\usepackage[english]{babel}

\usepackage{textcomp}

\usepackage{array}
\newcolumntype{L}[1]{>{\raggedright\let\newline\\\arraybackslash\hspace{0pt}}m{#1}}
\newcolumntype{C}[1]{>{\centering\let\newline\\\arraybackslash\hspace{0pt}}m{#1}}
\newcolumntype{R}[1]{>{\raggedleft\let\newline\\\arraybackslash\hspace{0pt}}m{#1}}

\begin{document}

\title{Spin-orbital model of stoichiometric LaMnO$_3$ with tetragonal distortions}

\author     {Mateusz Snamina}
\affiliation{Kazimierz Gumi\'nski Department of Theoretical Chemistry, Faculty of Chemistry,\\
             Jagiellonian University, Gronostajowa 2, PL-30387 Krak\'ow, Poland }

\author{Andrzej M. Ole\'s}
\affiliation{Marian Smoluchowski Institute of Physics, Faculty of Physics, Astronomy,
             and Applied Computer Science,\\ Jagiellonian
             University, Prof. S. \L{}ojasiewicza 11, PL-30348 Krak\'ow, Poland}
\affiliation{Max Planck Institute for Solid State Research,
             Heisenbergstrasse 1, D-70569 Stuttgart, Germany}

\date{\today}

\begin{abstract}
The model developed for LaMnO$_3$ addresses the spin-orbital order
by superexchange and Jahn-Teller orbital interactions in the cubic
(perovskite) symmetry up to now whereas real crystal structure is
strongly deformed. We identify and explain three \textit{a priori}
important physical effects arising from tetragonal deformation:
(i) the splitting of $e_g$ orbitals $\propto E_z$,
(ii) the directional renormalization of $d-p$ hybridization $t_{pd}$,
and
(iii) the directional renormalization of charge excitation energies.
Using the example of LaMnO$_3$ crystal we evaluate their magnitude. It
is found that the major effects of deformation are enhanced amplitude
of $x^2-y^2$ orbitals induced in the orbital order by $E_z\simeq 300$
meV and anisotropic $t_{pd}\simeq 2.0$ (2.35) eV along the $ab$ ($c$)
cubic axis, in very good agreement with the Harrison's law.
We show that the tetragonal model analyzed within mean field
approximation provides a surprisingly consistent picture of the ground
state. Excellent agreement with the experimental data is obtained
simultaneously for:
(i) $e_g$ orbital mixing angle,
(ii)~spin exchange constants, and
(iii) the temperatures of spin and orbital phase transition.
\end{abstract}


\maketitle

\section{Introduction}

Manganites, cuprates and vanadates are wide groups of compounds that
have been attracting much attention, both theoretical and experimental.
What makes them intriguing are strong electron correlations that cause
electron localization \cite{Ima98} and thus lead to extremely complex
quantum behavior of coexisting spin and orbital degrees of freedom
\cite{Kug82}.
Theoretical description is even more demanding due to structural phase
transitions. To capture low energy phenomena at strong correlations,
superexchange models were introduced. Then these models were extended
so that they take a Jahn-Teller (JT) effect \cite{Mil96,Vol10} into
account.
The models may employ the parameters derived from \textit{ab initio}
calculations or from experiment and thus serve as satisfactory and
convincing explanation of interactions in the space of spin-orbital
degrees of freedom \cite{Kha05,Ole05}. Several nontrivial disordered
\cite{Rei05,Mila} or ordered \cite{Karlo} phases arise as a generic
consequence of spin-orbital exchange interactions, including novel
phases found by orbital \cite{Brz15} or charge \cite{Brz16} dilution.
Though such models have been successful in working out the peculiar
physical properties of doped manganites \cite{Dag01,Tok06},
the structural aspects have usually been neglected. One of them is
the interplay between a crystal geometry and the actual spin-orbital
interactions.

One of the physical aspects of perovskites that has not yet been fully
investigated is a shortening of some bonds in the crystal structure
and elongation of the others. Any deformation of the crystal results
in the lowering of lattice symmetry from cubic to tetragonal. In this
paper we are peering at this structural aspect in details. We choose
LaMnO$_3$ as a guide compound as it is described in many textbooks and
the spin-orbital superexchange model is well known \cite{Fei99} and
has been used to explain the temperature dependence of the spectral
weights observed in the optical spectroscopy \cite{Kov10}. Hund's
exchange stabilizes large spins $S=2$ in LaMnO$_3$ and quantum effects
are then reduced. Thus spin-orbital entanglement is small as we have
shown in a recent study \cite{Snm16}. Then an additional advantage is
that the analysis is simpler here than for some systems with smaller
spins (such as $S=1/2$ in KCuF$_3$) where spin-orbital entanglement
cannot be neglected \cite{Ole12,You15}. However, superexchange alone
is not sufficient to explain the high value of the orbital transition
temperature $T_{\rm OO}$ \cite{Fei98,Fei99,Oka02,Pav10}. A careful
study of the orbital melting transition suggests that superexchange
interactions play a minor role for this transition while tetragonal
crystal-field (CF) splitting has to be included to explain experiments
\cite{Fle12}. This is an argument to go beyond the cubic model
of LaMnO$_3$.

Indeed, the real LaMnO$_3$ crystal structure is much more complicated
than that of an ideal cubic perovskite. It may be perceived as a cubic
perovskite with one period, i.e., bond length in antiferromagnetic (AF)
$c$ direction is significantly shorter than the ferromagnetic (FM)
bonds along $a$ and $b$ axes (with the difference between them of about
3.5\% of the initial length). This tetragonal deformation is thus
\textit{opposite} to that found in high-$T_c$ cuprates, where the
apical oxygens are more distant from $3d$ ions than the oxygens in the
$ab$ planes, as predicted by the electronic structure calculations
\cite{Saw97} and confirmed experimentally \cite{Crv98}. In addition,
MnO$_6$ octahedra are slightly tilted (so that the space group is
\textit{Pmna}). In this work we modify the cubic model so that it takes
into account the differences of period lengths but completely neglects
octahedra tilts. We show below that the period lengths difference is
related to the orbital state. The bigger the lengths difference is, the
more robust the orbital state becomes. For this reason the onset of an
orbital order takes place simultaneously with a structural phase
transition \cite{Crv98,Miz99,Alo00}.

In our previous work \cite{Snm16} we concentrated on the influence
of the diverse entanglements (spin-orbital on-site and on-bond
entanglement) on the electronic state of the LaMnO$_3$ crystal. The
goal of the present work is to assess the influence of the crystal
tetragonal deformation and to identify the main underlying physical
mechanisms playing a role for a realistic description. As the
LaMnO$_3$ crystal is a representative compound that is described by
Kugel-Khomskii-like model that undergoes strong JT effect,
we argue that the present considerations may be treated as a
guideline for any similar model extensions for analogous crystals.

Key degrees of freedom of the LaMnO$_3$ crystal are associated with
manganese ions. They have $3d^4$ high-spin (HS) $t_{2g}^3e_g^1$
configuration with spin $S=2$. Thus each manganese ion has large
magnetic moment that may point at any direction (due to SU(2)
symmetry). In addition to the magnetic degree of freedom there is an
orbital one due to a single $e_g$ electron. As every manganese ion is
at a center of (slightly deformed) MnO$_6$ octahedron, its $3d$
orbitals are split and the HS $d^4$ configuration includes the occupied
$e_g$ orbital with lower energy. The $e_g$~electron may occupy one of
the two basis orbitals (labeled in analogy to $|{\uparrow}\rangle$ and
$|{\downarrow}\rangle$ spin states):
\begin{equation}
\label{real}
\textstyle{
|\zeta_c\rangle\equiv \frac{1}{\sqrt{6}}(3z^2-r^2),
\hspace{0.7cm}
|\xi_c\rangle\equiv \frac{1}{\sqrt{2}}(x^2-y^2),}
\end{equation}
so that orbital state at site $i$ is, in general, a linear combination
of two basis states \cite{Kho14},
\begin{equation}
 |i\vartheta\rangle =
     \cos(\vartheta/2) \big|i\zeta_c\big\rangle
   + \sin(\vartheta/2) \big|i\xi_c\big\rangle.
\label{mixing}
\end{equation}
The orbital basis states in Eq. (\ref{real}) are obtained for
$\vartheta=0$ and $\vartheta=\pi$, respectively, while for angles
increased by $2\pi/3$ or $4\pi/3$ two other equivalent pairs of states
are obtained: $\{\zeta_a,\xi_a\}$ and $\{\zeta_b,\xi_b\}$.
The orbital state at site $i$, $|i\vartheta\rangle$, is parametrized by
\textit{orbital mixing angle} $\vartheta\in [0,2\pi)$. Thus the state
at each manganese ion is described by the direction of its spin
projection and the angle $\vartheta$. A 3D cubic lattice of manganese
ions in the LaMnO$_3$ crystal may just be viewed as a set consisting of
the above pairs of variables, representing the degrees of freedom of
each ion in the lattice.

The spin order in the ground state of LaMnO$_3$ crystal is $A$-type AF
(\mbox{$A$-AF}). Is means that the crystal is made of FM $ab$ planes
that are staggered in an AF manner along the $c$ axis. The orbital
state is nontrivial as well. The manganese's $e_g$ electron states
are equal to $|\pm\vartheta\rangle$, where the angle
$\vartheta\in[0,\pi)$ takes for some fixed value and the sign
alternates between the $A$ and $B$ sublattice in each $ab$ plane
\cite{Ole05}. The orbital state is unchanged along the perpendicular
$c$ axis, i.e., the alternating orbital (AO) order is $C$-type labeled
as \mbox{$C$-AO}.

The purpose of this paper is to introduce and investigate the
consequences of tetragonal crystal structure of LaMnO$_3$ which takes
into account the experimental distortions. It is noteworthy that the
cubic model predicts the reduction of symmetry in the ordered state
as one direction is distinguished by the spin-orbital order
(the AF direction in \mbox{$A$-AF}/\mbox{$C$-AO} state). In this state
the symmetry is broken, i.e., the ground state has lower symmetry
than the symmetry of the model itself. On the contrary, the proposed
tetragonal model has lower symmetry from the beginning.
Following the observed structure \cite{Hua97}, in this work we label
the ``shortened'' AF directions by letter $c$, and the ``elongated''
FM directions $a$ and $b$.

The paper is organized as follows. In Sec. \ref{sec:model} we develop
spin-orbital model for LaMnO$_3$ in the tetragonal phase. We begin
with recalling the model for the usually considered cubic structure in
Sec. \ref{sec:cubic}, present the tetragonal structure in Sec.
\ref{sec:struc}, and summarize the necessary changes in Sec.
\ref{sec:changes}. In Sec. \ref{sec:tetra} we concentrate on the
tetragonal crystal field (CF):
(i) introduce its microscopic description in Sec. \ref{sec:micro},
(ii) present its consequences on the ground state in Sec.
\ref{sec:imcf}, and
(iii) determine the actual value of the $e_g$ orbital splitting by an
\textit{ab initio} approach in Sec. \ref{sec:abi}.
The lattice distortions in LaMnO$_3$ lead to the renormalized
superexchange model presented in Sec. \ref{sec:reno}. We begin with
recalling the perturbative origin of the spin-orbital model in Sec.
\ref{sec:origin} and then discuss the renormalization of both
hybridization $t_{pd}$ hopping elements and charge excitation energies
in Secs. \ref{sec:tpd} and \ref{sec:exci}.
The predictions of the tetragonal model at $T=0$ are given in Sec.
\ref{sec:T=0}, and next discussed in Sec. \ref{sec:dis}.
As the tetragonal distortion changes together with electronic state,
the predictions of the model at finite temperature are distinct from
those of the cubic model as we show in the Appendix.
Finally we present the main conclusions and summarize the present
study in Sec. \ref{sec:summa}.

\section{Tetragonal model for L\lowercase{a}M\lowercase{n}O$_3$}
\label{sec:model}

\subsection{Spin-orbital model for cubic L\lowercase{a}M\lowercase{n}O$_3$}
\label{sec:cubic}

In the previous works the LaMnO$_3$ crystal electronic structure was
investigated with the aid of pure electronic superexchange
Kugel-Khomskii-like models \cite{Fei99,Kov10,Snm16}. Typically, such
models are formulated in terms of the ionic spin-orbital degrees of
freedom for Mn ions (all other degrees of freedom are integrated out,
including degrees of freedom attributed to the bridge atoms).
All terms in the superexchange Hamiltonian correspond to the ionic
pairs on nearest neighbor bonds (denoted here as $\langle ij\rangle$).

Charge excitations responsible for superexchange arise from electron
hopping $t$ defined for $e_g$ electrons as the largest hopping element
for a $\sigma$-bond $\langle ij\rangle\parallel\gamma$, i.e., between
two active orbitals, $|i\zeta_{\gamma}\rangle$ and
$|j\zeta_{\gamma}\rangle$ along this bond. In the cubic crystal $t$ is
independent of the bond direction $\gamma$. In the regime of large
intraorbital Coulomb repulsion $U\gg t$, one can construct the
low-energy Hamiltonian by attributing each virtual excitation numerated
by subscript $n$ with the Hamiltonian contribution,
\beq
\label{sex}
H_n^{\gamma}(ij) = \left( a_n + b_n \vec S_i \cdot \vec S_j \right)
{\cal Q}_n^{\gamma}\left(ij\right),
\eeq
where $a_n$ and $b_n$ are numeric coefficients derived from the
multiband extended Hubbard model; the explicit form of the Hamiltonian
was presented in Ref. \cite{Snm16}. Here
$\{\vec S_i\}$ denotes the spin operator (for $i$'th site) and
$Q_n^{\gamma}(ij)$ denotes the on-bond orbital operator, specifying the
orbital configuration. It is expressed in terms of on-site orbital
operators $\big\{\tau_i^{(\gamma)}\big\}$
(that explicitly depend on the crystallographic direction $\gamma$
of the $\langle ij\rangle$ bond):
\bea
\label{orbop}
\tau^{(a)}_i&=& -\frac14\sigma^z_i-\frac{\sqrt{3}}{4}\sigma^x_i, \nonumber \\
\tau^{(b)}_i&=& -\frac14\sigma^z_i+\frac{\sqrt{3}}{4}\sigma^x_i, \nonumber \\
\tau^{(c)}_i&=&  \frac12\sigma^z_i,
\eea
where $\sigma^z_i$ and $\sigma^x_i$ are Pauli matrices acting in the
space spanned by the orbital basis Eq. (\ref{real}) at site $i$.

A charge excitation between two transition metal ions with partly filled
$e_g$-orbitals will arise by a hopping process between two active
orbitals, $|i\zeta_{\gamma}\rangle$ and $|j\zeta_{\gamma}\rangle$.
To capture the directional dependence of such processes we introduce two
projection operators on the orbital states for each bond,
\begin{eqnarray}
\label{porbit}
{\cal Q}_\perp^{(\gamma)}(ij)&\equiv&
2\left(\frac14-\tau^{(\gamma)}_i\tau^{(\gamma)}_j\right),  \\
\label{qorbit}
{\cal Q}_\parallel^{(\gamma)}(ij)&\equiv&
2\left(\frac12+\tau^{(\gamma)}_i\right)\left(\frac12+\tau^{(\gamma)}_j\right).
\end{eqnarray}
Unlike for a spin system, the charge excitation
$d_i^md_j^m\rightleftharpoons d_i^{m+1}d_j^{m-1}$ is allowed only in
one direction when one orbital is directional $|\zeta_{\gamma}\rangle$
and the other is planar $|\xi_{\gamma}\rangle$ for a given bond
$\langle ij\rangle\parallel\gamma$, i.e.,
$\left\langle{\cal Q}_\perp^{(\gamma)}(ij)\right\rangle=1$;
such processes generate both HS and low-spin (LS) contributions.
On the contrary, when both orbitals are directional, i.e., one has
$\left\langle{\cal Q}_\parallel^{(\gamma)}(ij)\right\rangle=2$,
only LS terms contribute.

Now, it it straightforward to write the formula for the low-energy
spin-orbital Hamiltonian \cite{Snm16} which includes the superexchange
terms due to $e_g$ ($H_J^e$) and $t_{2g}$ ($H_J^t$) charge excitations,
the JT orbital interactions ($H_{\rm JT}$), and tetragonal crystal
field ($H_z$),
\begin{equation}
{\cal H}_{\rm som} = H_J^e + H_J^t + H_{\rm JT} + H_z,
\label{som}
\end{equation}
where
\begin{align}
\label{He}
\begin{split}
 H_J^e = J\!\sum_{\langle ij \rangle\parallel\gamma}
&\left\{-\frac{1}{40} r_1
      \left( \vec S_i\cdot\vec S_j+6 \right) {\cal Q}_\perp^{(\gamma)}(ij)\right. \\
    &+\left.\frac{1}{320}\left( 3 r_2 + 5 r_3 \right)
      \left( \vec S_i\cdot\vec S_j-4 \right) {\cal Q}_\perp^{(\gamma)}(ij)\right. \\
    &+\left.\frac{1}{64}\left( r_4 +  r_5 \right)
      \left( \vec S_i\cdot\vec S_j-4 \right) {\cal Q}_\parallel^{(\gamma)}(ij)\right\},
\end{split}
\end{align}
and
\begin{equation}
\label{Ht}
 H_J^t = \frac19\, J\sum_{\langle ij \rangle}
         r_t \left( \vec S_i \cdot \vec S_j - 4 \right).
\end{equation}
The superexchange energy $J=4t^2/U$ is here isotropic. The multiplet
structure of $e_g$ excited states is given by $\eta_e\equiv J_H^e/U$
which defines the coefficients,
\begin{eqnarray}
 r_1&=& \frac{1}{1-3\eta_e}, \hskip 1.8cm
 r_2 = \frac{1}{1+3\eta_e/4},  \nonumber \\
 r_3&=& r_4 = \frac{1}{1+5\eta_e/4}, \qquad
 r_5 = \frac{1}{1+13\eta_e/4}.
\end{eqnarray}
For $t_{2g}$ charge excitations are given by $\eta_t\equiv J_H^t/U$
and it is convenient to introduce a single coefficient \cite{Snm16},
\begin{equation}
r_t\! =
 \frac{1}{8}\left( \frac{1}{4+5\eta_t}+ \frac{1}{4+9\eta_t}
+\frac{1}{4+11\eta_t}+ \frac{1}{4+15\eta_t} \right)\!.
\end{equation}

The JT Hamiltonian $H_{\rm JT}$ describes the coupling between the
adjacent sites via the mutual octahedron distortion. We note that the
JT effect is connected with the oxygen atoms displacements which result
in longer and shorter Mn-O bonds in $ab$ planes but leave the manganese
positions unchanged. The JT term is controlled by a single parameter
$\kappa$ that describes the rigidity of the oxygen positions (or the
magnitude of displacement caused by the adjacent Mn orbital state).
If the oxygens were rigid and their positions were not influenced by
manganese orbital states then $\kappa=0$, but in reality $\kappa>0$.
The effective orbital intersite interaction term is given by the
orbital operators $\big\{\tau_i^{(\gamma)}\big\}$ as follows,
\begin{equation}
 H_{\rm JT} =
 8\kappa \sum_{\langle ij\rangle } \tau_i^{(\gamma)} \tau_j^{(\gamma)}.
 \label{HJT}
\end{equation}
This term favors AO order in the $ab$ FM planes and dominates over the
superexchange \cite{Fle12,Dag04}. Finally, the last term in Eq.
(\ref{som}) stands for the tetragonal splitting of $e_g$ orbitals and
is introduced below in Sec. \ref{sec:tetra}. Both $H_{\rm JT}$ and
$H_z$ (see Sec. \ref{sec:micro}) modify the $e_g$ superexchange via
the orbital order at zero temperature ($T=0$) \cite{Fei99}.

\begin{table}[b!]
\caption{Microscopic parameters of LaMnO$_3$ defining the spin-orbital
cubic model of Ref. \cite{Snm16} (all in eV):
effective $(dd\sigma)$ hopping $t$,
intraorbital Coulomb element $U$,
Hund's exchange for $e_g$ ($J_H^e$) and $t_{2g}$ ($J_H^t$) electrons,
and the orbital-orbital interaction induced by the JT effect
$\kappa$ (\ref{HJT}).}
\begin{ruledtabular}
\begin{tabular}{ccccc}
  $t$  &  $U$ & $J_H^e$ & $J_H^t$ & $\kappa$ \cr
 \hline
 0.37  &  4.0 &   0.69  &   0.59  &  0.006   \cr
 \end{tabular}
\end{ruledtabular}
\label{tab:para}
\end{table}

In our previous paper \cite{Snm16} we investigated the consequences
of the model of the LaMnO$_3$ crystal described in the literature,
using the established parameters listed in Table \ref{tab:para}. The
tetragonal splitting of $e_g$ orbitals $\propto E_z$ was neglected.
We performed mean field (MF) (cluster) calculations at finite
temperatures. This allowed us to fit the values of parameters in the
spin-orbital model, to make it the most reliable for the experimental
situation. Our scrutiny revealed however, that even in that case, the
model predictions are not fully consistent with the experiments
(especially, when it comes to the value of the orbital mixing angle).
We eliminated the possibility that the discrepancy may stem from the
way we solved the model. (We showed that the short-range correlations
play only moderate role in this case and cannot break the MF-based
methodology). This prompted us to search for the origin of the
discrepancy in the model main assumptions,
especially the ones implementing the model cubic symmetry.

\subsection{Tetragonal geometry of LaMnO$_3$}
\label{sec:struc}

In the LaMnO$_3$ crystal the manganese ions are arranged in a tetragonal
lattice. In terms of the \textit{Pmna} space group crystallographic axes
$a$, $b$, and $c$, the \mbox{Mn-Mn} distances are given by
$\frac12\sqrt{a^2+b^2}$ in the FM $ab$ planes and $c/2$ along the AF
$c$ axis. The measured (at $T=300$~K) values of these structural
constants are \cite{Hua97}: $a=5.5378$\AA{}, $b=5.7385$\AA{}, and
$c=7.7024$\AA{}, so that the \mbox{Mn-Mn} distances are equal to
$3.9874$\AA{} in the $ab$ planes and $3.8512$\AA{} along the $c$ axis.
The relative difference in the lengths is as big as 3.5\%.

To appreciate fully the consequences of tetragonal distortion, one has
to include oxygen bridge positions in between neighboring manganese
pairs which influence the electronic structure.
The exact values of the corresponding \mbox{Mn-O} distances depend on:
(i) the overall distance of two involved manganese ions and
(ii) JT-like oxygens displacements.
The JT effect is taken into account by specially designed effective JT
Hamiltonian Eq. (\ref{HJT}) with $\kappa\simeq 6$ meV,
see Table \ref{tab:para}, and there is no
need to change it on the first instance. On the other hand, to assess
the influence of the anisotropy of \mbox{Mn-Mn} distances, the cubic
model has to be somehow extended.

As the oxygens JT-like displacements are described separately, the
starting geometry for our tetragonal crystal model is established by
putting the oxygen atoms directly in the half-way between the
neighboring manganese ions. Thus the \mbox{Mn-O} bond lengths are set
to $1.9937$\AA{} and $1.9256$\AA{}, respectively.
In such a geometry there are deformed MnO$_6$ octahedra of $D_{4h}$
point symmetry. In this work we will present the results for "rounded"
values $1.995$\AA{} (for the planar $ab$ axes) and $1.925$\AA{}
(along the $c$ axis). These values may be compared with the value for
non-deformed MnO$_6$ octahedra with \mbox{Mn-O} distances equal to
$1.970$\AA{}.

\subsection{Changes in comparison to cubic model}
\label{sec:changes}

We identify three main modifications that should be made to transform
the well known cubic model \cite{Fei99} into a tetragonal model:
\begin{itemize}
\item{(i)}
The first change concerns \textit{tetragonal} CF effect itself. While
displacements of oxygen atoms are well described by the JT Hamiltonian,
the \mbox{Mn-Mn} distances influence the spin-orbital state by
\textit{tetragonal} CF. The \textit{tetragonal} CF was introduced in
Ref. \cite{Fei99} but its strength was not widely discussed.
\item
The second change concerns the $d-p$ hopping integral. 
In superexchange theory the virtual processes are perceived as
a sequence of electron hopping's. The closer involved atoms lie,
the stronger superexchange is.
As far as we know this effect has never been discussed.
\item
The third change concerns the energies of charge excitations
$\{\varepsilon_n\}$, given in Refs. \cite{Ole05,Kov10}. In the
superexchange theory excited virtual configurations are considered
(for example the ${}^6A_1$ term for the HS $d^5-d^3$ excitation). The
charge excitation energies depend on relative geometry (especially atom
distances). In the tetragonal model the interatomic distances in one
direction are different than the distances in the other two directions.
Hence the corresponding energies in the model should be renormalized.
As far as we know this effect also has not been discussed earlier.
\end{itemize}

In this work we assess the importance of all the above changes.
As necessary steps we evaluate the strength of the \textit{tetragonal}
CF and the change in the hopping integrals using remote
quantum-chemical models.

\section{Tetragonal crystal-field splitting}
\label{sec:tetra}

\subsection{Definition of the crystal-field splitting}
\label{sec:micro}

We investigated a generalization of the LaMnO$_3$ cubic model by
introducing tetragonal CF. The \textit{tetragonal} CF is modeled by
the Hamiltonian $H_z$ that is governed by a single parameter $E_z$,
describing the energetic splitting of the $e_g$ states (\ref{real}):
\begin{equation}
\label{Hz}
 H_z =\frac12 E_z\sum_i\left(|i\zeta_c\rangle\langle i\zeta_c|
                            -|i\xi_c\rangle\langle i\xi_c|\right)
     =        E_z\sum_i\tau_i^{(c)}.
\end{equation}
For positive values ($E_z>0$), $|\xi_c\rangle$ (i.e., $|x^2-y^2\rangle$
states) are favored, whereas for negative values ($E_z<0$),
$|\zeta_c\rangle$ (i.e., $|3z^2-r^2\rangle$) states have lower energy.
For the realistic structure of LaMnO$_3$ one has $E_z>0$. The bigger
the value $E_z$ is, the greater the contribution of $|x^2-y^2\rangle$
orbitals is obtained (and the value of $\vartheta$ increases approaching
$180^{\degree}$). If the \textit{tetragonal} CF is ignored ($E_z=0$)
then the orbital mixing angle (at $T=0$) is equal to $84^{\degree}$
\cite{Ole05} contradicting the experiment; the experimental value
${\vartheta_{\rm exp}=108^{\degree}}$ \cite{Crv98} is reproduced in the
model with $E_z \simeq 190$~meV (see the following Section).

\subsection{Impact of the CF splitting on spin-orbital order}
\label{sec:imcf}

We used on-site MF calculation scheme to determine the value of
orbital mixing angle~$\vartheta$, \textit{energy per site}~$E$,
and spin \textit{exchange constants}~$J_\gamma$ ($\gamma=a,b,c$).
The calculations revealed that the $A$-AF/$C$-AO state is destroyed by
sufficiently strong tetragonal CF and the way the pattern decomposes
under the influence of the tetragonal CF is quite non-trivial.

The spin $A$-AF pattern is obtained only when spin exchange constants
fulfill $J_a=J_b<0$ and $J_c>0$. Their values depend on $\vartheta$
which, in turn, strongly depends on the actual CF splitting~$E_z$
(\ref{Hz}). It turns out that the first constraint ($J_a=J_b<0$) is not
valid for ${\vartheta>\vartheta^\text{cr}\simeq 122^\degree}$ that
occurs when the $A$-AF state breaks down at $E_z^\text{cr}\simeq 291$~meV.

However, in reality, even when tetragonal CF is weaker than the critical
value, $E_z<E_z^\text{cr}$, the $A$-AF state may not be the true ground
state. Calculations reveal that even comparatively weak tetragonal CF
may be strong enough to trigger spin reorientation. More precisely:
When tetragonal CF is sufficiently strong, it is favorable for the spins
to turn around and to change their order discontinuously from $A$-AF
into $G$-AF. The transition is accompanied by discontinuous jump of the
\textit{orbital mixing angle} (from $\sim 110^{\degree}$ to
$\sim 150^{\degree}$).
The transition occurs for $E_z=E_z^\text{cr}\simeq 188$~meV.

\begin{figure}[t!]
 \begin{center}
 \includegraphics[width=\columnwidth]{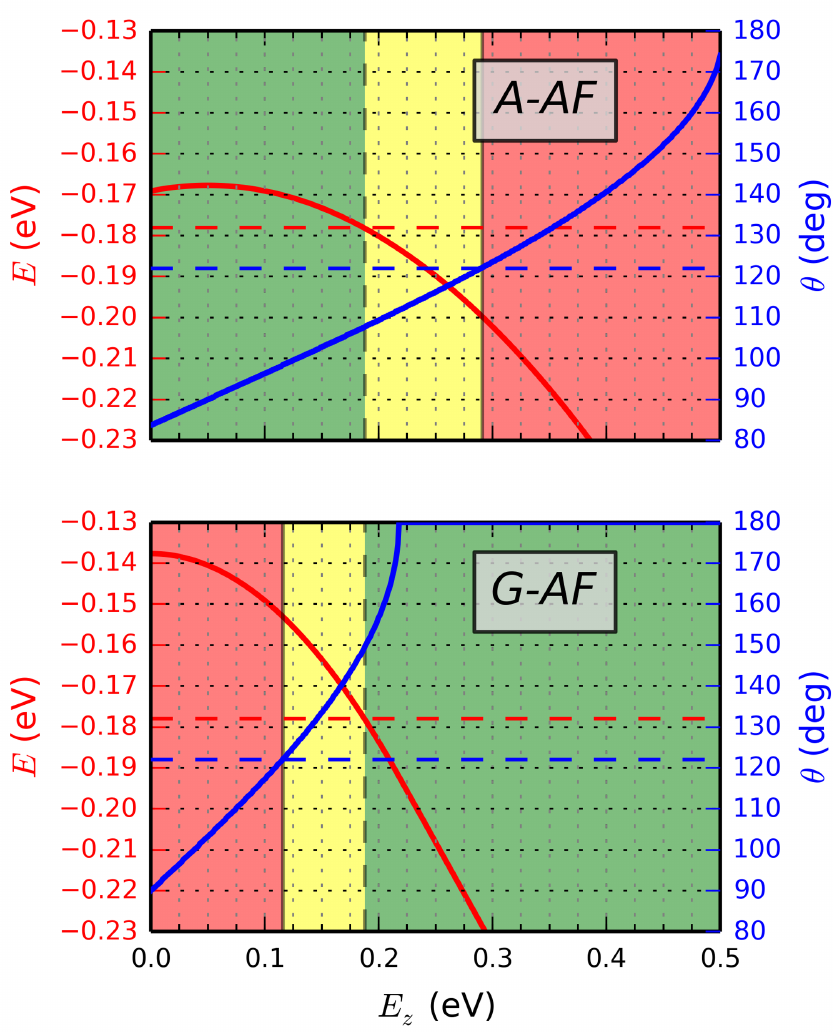}
 \end{center}
\caption{Graphical representation of the influence of the tetragonal
CF $E_z$ (\ref{Hz}) on the:
$A$-AF state (upper panel) and
$G$-AF state (lower panel).
Other parameters as in Table \ref{tab:para}.
Background colors indicate the regimes of:
green  --- the ground state,
yellow --- a metastable excited state, and
red    --- inconsistency for each type of spin order.
Solid lines show:
the \textit{energy per site} $E$ (red) and
the \textit{orbital mixing angle} $\vartheta$ (blue).
The horizontal dashed red line indicates the energy of
the configurations at the quantum phase transition (at $E_z^\text{QPT}$),
while the horizontal dashed blue line indicates  $\vartheta^\text{cr}$.}
\label{fig:EzInpact}
\end{figure}

A physical explanation of the transition between different magnetic
states is as follows: The $A$-AF state coexists with the orbital state
characterized by ${\vartheta_\text{($A$-AF)}\simeq 110^\degree}$,
whereas $G$-AF state coexists with the orbital state characterized by
${\vartheta_\text{($G$-AF)}\simeq 150^\degree}$.
(Given values are for $E_z \simeq 200$~meV; of course
$\vartheta{\nearrow}$ when $E_z{\nearrow}$ for the both cases but a key
relation ${\vartheta_\text{($G$-AF)} \gg \vartheta_\text{($A$-AF)}}$ is
valid for wide range of $E_z$.) In the absence of tetragonal CF,
the $A$-AF state is the ground state. On the other hand, the tetragonal
CF prefers states with larger angle~$\vartheta$. As one state is
preferred when tetragonal CF is weak and the other is favored at strong
tetragonal CF, the transition in between is rationalized.

The results of our calculations are summarized in Fig.~\ref{fig:EzInpact}.
For the both types of spin order ($A$-AF and $G$-AF) their regimes of
existence are indicated (background color different than red).
Then, each such regime of existence is divided into two parts where the
considered type of spin order is:
the ground state (green background color) and
the excited state (yellow background color).
In addition, for the both spin states \textit{energy per site} $E$
and \textit{orbital mixing angle} $\vartheta$ are plotted
(as function of $E_z$).

It is important to mention that in on-site MF calculation scheme
a $C$-AO state and a $G$-AO state (both with the same spin order)
give rise to solutions with identical \textit{energy per site} $E$
and \textit{orbital mixing angle} $\vartheta$. These states may be
perceived as stacks of planes of the same type
(namely: anti-ferro-orbital planes).
What differs the states is a relative displacement of the adjacent
planes: the inter-plane bonds are either
${|\pm\vartheta\rangle|\pm\vartheta\rangle}$ (in $C$-AO phase), or
${|\pm\vartheta\rangle|\mp\vartheta\rangle}$ (in $G$-AO phase).
The energy contributions from ``in-plane bonds''
(bonds~$\langle ij\rangle{\parallel}ab$) are the same for both
states (as they consist of the same planes) and do not differentiate
the two states. Each energy contribution from ``inter-plane bond''
(bond~$\langle ij \rangle{\parallel}c$) depends on the orbital
operators $\{\tau^{(c)}_i,\tau^{(c)}_j\}$.
However, exclusively for direction $c$:
$\big\langle + \vartheta\big|\tau^{(c)}\big| + \vartheta\big\rangle
=\big\langle - \vartheta\big|\tau^{(c)}\big| - \vartheta\big\rangle$,
so that the prospective structural difference has no influence on the
bonds energy contributions. As a result the adjacent planes relative
orientation does not matter for energy value.
The results presented in this chapter are for $C$-AO and $G$-AO pattern;
we have verified that solutions for ferro-orbital (FO)
or $A$-AO patterns have higher energy.

\subsection{Ab-initio evaluation of tetragonal CF strength}
\label{sec:abi}

The strength of the tetragonal CF splitting $E_z$ (\ref{Hz})
distinguishes between two HS states:
$t_{2g}^3z^1$~state and $t_{2g}^3\bar{z}^1$~state.
The splitting depends on the actual MnO$_6$ octahedron
deformation (in other words: on \mbox{Mn-O} bond lengths).
For calculations we take the reference geometry in which
\mbox{Mn-O}${\parallel}c$  axis bond length is equal to $1.925$\AA{} and
\mbox{Mn-O}${\parallel}ab$ plane bond length is equal to $1.995$\AA{}.
The deformed octahedron anion [Mn${}^{III}$O${}_6$]$^{9-}$ was used
as a quantum chemical model. The anion was immersed in charge lattice
that is to mimic the other ions that made up the crystal. The final
splitting values obtained in different quantum chemical levels of
theory are shown in Table \ref{tab:splittingEz}.

At the beginning we carried out Hartree-Fock (HF) calculations in
minimal basis set (STO-3G) in given geometry and imposed multiplicity
(quintet). We used a standard linear combination of atomic orbitals
(LCAO) scheme of unrestricted HF implementation (UHF). Each molecular
HF calculation run was followed by the standard promolecule calculation
that sets an initial-guess density matrix.
As expected on grounds of ligand field theory the obtained HF state's
highest occupied molecular orbital (HOMO) is $e_g^*$ molecular orbital
(MO) with significant occupancy of the $\bar{z}$ atomic orbital (orbital
symmetry: $B_{1g}$) at manganese ions whereas lowest unoccupied MO (LUMO)
is $e_g^*$ MO with significant occupancy of the $z^2$ atomic orbital
(orbital symmetry: $A_{1g}$). Hence, the obtained state corresponds to
the model $t_{2g}^3\xi_c^1$ configuration.

Then, the second HF calculation was carried out.
But in this case, the calculation started from a hand-made initial-guess
density matrix determined from MOs obtained in the previous HF
calculation run --- the initial-guess configuration was build of the
LUMO and all occupied MOs but the HOMO. As in course of HF iterations
MOs was changed only quantitatively (not qualitatively) the obtained
state was labeled as the model $t_{2g}^3z^1$ configuration.
A desired outcome of these calculations --- the strength of tetragonal
CF (parameter $E_z$) was calculated as a difference between the
energies of two obtained states. At the lowest level of the theory with
the minimal basis set (STO-3G) one finds $E_z^{(1)} = 469$~meV.

It is a common wisdom, that the calculations using the minimal basis
set give usually a decent physical insight, but are not of sufficient
quality and should be treated as pilot calculations only. Motivated by
that we attempted to repeat calculations using some bigger Popple basis
sets. Unfortunately the calculations were misleading. The first HF run
ended up with converged HF (ground) state made of MOs that bear no
resemblance to the standard ligand field theory like orbitals
(e.g. $e_g^*$ orbitals).
Thus the obtained energies could not be used in the evaluation of $E_z$.

To overcome this problem we designed a modified calculation strategy
for bigger basis sets. The modification applies to initial-guess states
(implemented for both HF calculation runs:
the one intended to describe $t_{2g}^3\bar{z}^1$ configuration and
the other one for            $t_{2g}^3z^1$     configuration).
The initial-guess states were calculated as projections of the
corresponding states obtained in the minimal basis set calculations.
The projections allow us to translate state description from one basis
set to another basis set. (Of course the translated state is not strictly
equivalent of the original state, but it is the closest possible match.)
Then the typical HF convergence run was carried out. For 6-31G basis set
calculations the HF iterations, fortunately, changed the initial guesses
only quantitatively and we arrived at the proper solutions for this
improved second level of theory.
The corresponding final result is $E_z^{(2)} = 309$~meV.

\begin{table}[t!]
\caption{The predicted values of tetragonal splitting $E_z$ (in meV)
in LaMnO$_3$ at different sophistication levels of the theory.}
\begin{ruledtabular}
\begin{tabular}{ccccc}
number & method & basis set & ext. charges & $E_z$  \\
 \hline
 1     &   HF   &  STO-3G   &  \checkmark  & $469$ \\
 2     &   HF   &  6-31G    &  \checkmark  & $309$ \\
 3     &   HF   &  6-311G   &  \checkmark  & $296$ \\
 4     &   HF   &  6-311G*  &  \checkmark  & $299$ \\
 5 & DFT (B3LYP)&  STO-3G   &  \checkmark  & $434$ \\
 6     &   HF   &  STO-3G   &    absent    & $465$ \\
\end{tabular}
\end{ruledtabular}
\label{tab:splittingEz}
\end{table}

The convergence to the proper (ligand field theory like) solutions is
a matter of luck. Typically HF iterations tend to change the nature of
MOs so that calculations end in another physical solution. To avoid
undesired configurations one may relay on imposing the symmetry types
of occupied MOs. In our case it would imply imposing the total numbers
of $B_{1g}$-type occupied MOs and the total number of $A_{1g}$-type
occupied MOs. However, typical implementations of this trick applies
for a maximal Abelian symmetry subgroup
(as they have solely non-degenerate irreducible representations that
are easy to handle). Deformed MnO$_6$ octahedron has $D_{4h}$ symmetry,
with a maximal Abelian subgroup $D_{2h}$. In terms of $D_{2h}$ group
the $x^2-y^2$-like orbitals and $z^2$-like orbitals have the same
symmetry, i.e., $A_g$ (there are no $B$ type representations in
$D_{2h}$ group). This rules out the possibility of applying the above
trick in our study.

We repeated the modified procedure to obtain the results for bigger
and bigger basis sets. We were changing the basis gradually, see the
calculations number 1--4 in Table \ref{tab:splittingEz}. States were
projected many times, so that every HF run (expect for the minimal
basis set calculations) started with projected state obtained in only
slightly smaller basis calculation. The numeric values of $E_z$ for
larger than minimal basis set do not differ significantly so no further
correction is needed to predict a limiting value for an infinite basis,
\begin{equation}
E_z^\text{HF} \simeq 300 {\rm meV}.
\label{Ez}
\end{equation}

To verify this prediction we compared the obtained HF results for the
minimal basis set STO-3G (number 1) with the corresponding calculations
in DFT theory (we used popular B3LYP functional \cite{Bec93,Ste94}),
see number 5 in Table \ref{tab:splittingEz}, to check whether the
electron correlation effects may be significant in our problem. The
comparison reveals that HF (number 1) tends to overestimate the value
of $E_z$ for about 7\%. Eventually, we verified that, to our surprise,
the external charges in the lattice have a negligible effect,
see number 6 in Table \ref{tab:splittingEz}.

\section{Revised superexchange model}
\label{sec:reno}

\subsection{The origin of the superexchange Hamiltonian}
\label{sec:origin}

Integrating-out the high-energy intermediate states obtained by charge
excitations gives rise to the effective low-energy interactions; each
excitation contributes to spin-orbital superexchange Hamiltonian (which
we use). In this way energies of the excited charge configurations
appear in the effective Hamiltonian and define its parameters.

Both Mott insulators and CT insulators are described in terms of
superexchange when electron correlations are strong.
In both cases the sets of contributing virtual processes are the same,
however their relative importance depends on the actual parameters
which control these processes. In general, the sequence of four hopping
processes appears along an M$-$O$-$M bond,
\begin{align}
\begin{split}
  M^m L^{n} M^m
  &\xrightarrow{(1)}
  M^{m+1} L^{n-1} M^m
  \\ &\xrightarrow{(2)}
  M^{m+1} L^n M^{m-1}
  \\ &\xrightarrow{(3)}
  M^{m+1} L^{n-1} M^m
  \\ &\xrightarrow{(4)}
  M^m L^n M^m
  \label{eq:processes}
\end{split}
\end{align}
and describes leading processes for Mott insulators
(in a term $M^{m_1}L^{n_1}M^{m_2}$ the left-most $M$-ion electron shell
corresponds on one magnetic cation, the $L$ orbital shell in between
(of an O ion) belongs to the bridge ligand and the right-most $M$-ion
electron shell is localized on the neighboring magnetic cation). These
processes involve a hopping integral $t_{pd}$ between Mn($3d_{z^2}$)
orbital and bridge oxygen O($2p_z$) orbital for a bond along the bond
in each step.

In particular cases the electronic configurations of intermediate
states are realized by various atomic terms.
For example, in case of simple ``toy-model'' of a Mott insulator
for metals and ligands with only single interacting orbital, the
structure of terms is trivial and the low-energy Hamiltonian reads
\cite{Zaa88}:
\begin{multline}
 H_{\langle i j\rangle}\!
  = \frac{4t^4_{pd}}{\Delta^2}
    \left( \frac{1}{U_d} + \frac{2}{2\Delta + U_p} \right)
    \vec S_i \cdot \vec S_j
    \simeq
    \frac{4t^4_{pd}}{\Delta^2U}
    \vec S_i \cdot \vec S_j,
 \label{eq:toy_model}
\end{multline}
with $U_d$ ($U_p$) being the intraorbital Coulomb elements for $3d$
($2p$) orbitals, $\epsilon_d$ ($\epsilon_p$) being the electron energy
in orbital $3d$ ($2p$), and the charge transfer (CT) excitation energy
is
\begin{equation}
\Delta\equiv U_d-U_p+\epsilon_d-\epsilon_p.
\end{equation}
The first (second) term in Eq. (\ref{eq:toy_model}) is called Anderson
(Goodenough) contribution, both defined by the nature of charge
excitations, either at M or at O ion. Here we introduced an effective
Coulomb element $U$ to describe the superexchange by the Anderson
process. Note that $U\simeq U_d$ for Mott insulators, with
$\Delta\gg U_d$, while otherwise $U\simeq\Delta$.

Although the manganite case is much more involved than the "toy-model",
the general patterns of superexchange terms are similar in both cases.
In case of the LaMnO$_3$ crystal the values $U_d$ and $\Delta$ in Eq.
(\ref{eq:toy_model}) correspond to excitation energies to
${(3d)^5(2p)^6(3d)^3}$ and ${(3d)^5(2p)^5(3d)^4}$ configurations.
Hence, a general bond term in superexchange Hamiltonian is
proportional to:
\begin{multline}
 \bigg( \frac{t_{pd}^2}{E\big[(3d)^5(2p)^5(3d)^4\big]} \bigg)^2
 \frac{1}{E\big[(3d)^5(2p)^6(3d)^3\big]}\\
 \times ~
 \begin{pmatrix}
  \text{certain spin}\\
  \text{operator}
 \end{pmatrix}
 ~ \times ~
 \begin{pmatrix}
  \text{orbital projection}\\
  \text{operator}
 \end{pmatrix} .
 \label{eq:term}
\end{multline}
The expression in the first bracket describes the effective Mn$-$Mn
hopping,
\begin{equation}
t_\gamma\equiv \frac{t_{pd}^2(d_\gamma)}{E\big[(3d)^5(2p)^5(3d)^4\big]},
\label{tpd2}
\end{equation}
in the effective model ${\cal H}_{\rm som}$ (in which oxygen ions are
absent). Here $t_{pd}(d_\gamma)$ is the hybridization element for the
Mn$-$O bond of length $d_\gamma$. The exact form of the Hamiltonian was
derived in Ref. \cite{Fei99}. Our key observation is that the
parameters in Eq. \eqref{eq:term} depend on crystal geometry. Moreover,
the directional renormalization of $t_{pd}$ is amplified by a factor
of four in superexchange which results from processes of the type
$\propto t_\gamma^2/U$ in perturbation theory. In the two following
subsections we analyze separately the dependence on geometry of:
(i) $t_{pd}(d_\gamma)$ and
(ii) the energies of excited state.

\subsection{Renormalization of $t_{pd}$}
\label{sec:tpd}

We tried to assess the relative change of the value of hopping integral
$t_{pd}$ that follows from the \mbox{Mn$-$O$-$Mn} bonds length change.
Once more we built a quantum chemical model that serves to describe the
localized orbitals in the crystal. We believe that it is sufficiently
accurate to investigate the relative change in $t_{pd}$ value.

Is is quite nontrivial to assign concrete orbitals that correspond to
"metal $3d_{z^2}$ orbital" and "oxygen $2p_z$ orbital" --- the basis
orbitals in terms of elementary (not effective) superexchange models
(with metal and oxygen ions treated individually).
It is a fundamental assumption in these models that the orbitals are
orthogonal. It rules out the possibility of direct usage of the
corresponding atomic orbitals. It is a temptation to use MOs instead
(similar to that described in previous section). However this approach
is also not valid. In terms of the elementary superexchange model, for
which $t_{pd}$ is an parameter, the MOs are linear combinations of the
basis orbitals --- the results of coupling the basis orbitals mediated
by $t_{pd}$ (being an off-diagonal Hamiltonian matrix element).

\begin{figure}[t!]
 \centering
 \includegraphics[]{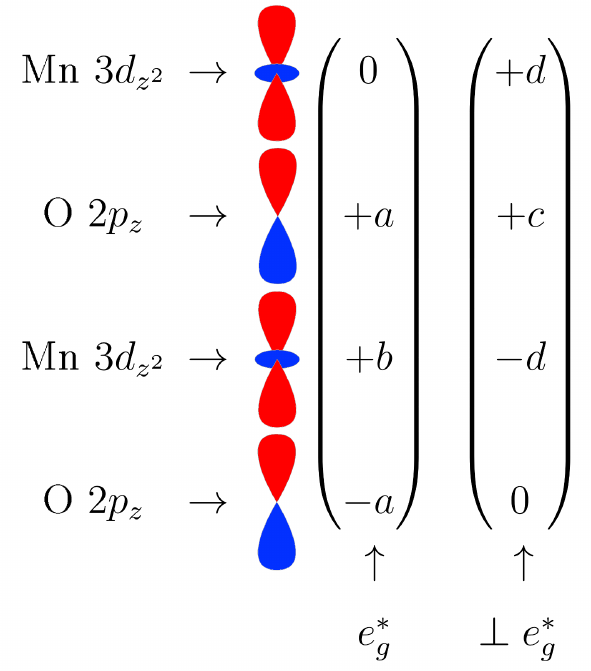}
\caption{The general form of basis MOs used to derive the superexchange
model. The values of $t_{pd}$ depend on the overlap between the adjacent
atomic orbitals. The parameters $b$ and $c$ are close to unity for the
orbitals centered at Mn and O ions along the $c$ axis, whereas the
values of $a$ and $d$ are close to zero (as long as electrons localize
and the overlap is small).}
 \label{fig:PairOfOrbitals}
\end{figure}

Here we propose to start with atomic orbitals (using STO-6G) and take
advantage of orthogonalization scheme that minimizes the orbital
changes, namely the L\"owdin orthogonalization. In this way we obtain
a pair of the orthogonal adjacent L\"owdin orbitals:
the metal $3d_{z^2}$-like orbital (labeled ``$e_g^*$'') and
the oxygen $2p_z$-like orbital (labeled ``${\perp}e_g^*$'')
of the form that is shown in Fig.~$\ref{fig:PairOfOrbitals}$.

Superexchange models introduce simple and intuitive descriptions in
which only a few electrons are treated explicitly, whereas the majority
of all remaining electrons (those occupying inner shells and some outer
shells) are treated implicitly. The latter are assumed to be fixed and,
together with nuclei, build up a specific scene in which the explicitly
treated electrons are immersed. In our case atoms configurations made
up by the implicitly treated electrons read:
\begin{displaymath}
  \text{Mn}^{4+}   :  \text{[Ar]} (3t_{2g})^3 ,~
  \text{O}(\gamma) :  \text{[He]} (2s)^2 (2p_{{\perp}\gamma})^4 ,~
  \text{La}^{3+}   :  \text{[Rd]}
\end{displaymath}
where $O(\gamma)$ denotes the oxygen atoms belonging to the Mn$-$O$-$Mn
bridges parallel to $\gamma$ direction and $2p_{{\perp}\gamma}$ denotes
pairs of $2p$ orbitals perpendicular to the $\gamma$ direction. We will
use L\"owdin orbitals (STO-6G) as a specific realization of the abstract
orbitals included in the above configuration scheme. This allows us to
use standard MF approximation to introduce interactions with the given
implicitly treated electrons.

The $t_{pd}$ is an one-electron mean-field Hamiltonian off-diagonal
matrix element. In terms introduced above one finds,
\begin{equation}
 \label{eq:t_pd_definition}
 t_{pd}
 =
 \Big\langle
   e_g^*
 \Big|
   H_{1e}^{\rm MF}
 \Big |
   {\perp}e_g^*
 \Big\rangle,
\end{equation}
where
\begin{equation}
 \label{eq:hamiltonian_1e_mf}
 \hat H_{1e}^{\rm MF}
 =
 \hat T + \hat V_{en} + \hat J + \hat K,
\end{equation}
encompasses: a single electron kinetic energy ($\hat T$),
electron-nuclei Coulomb interaction ($\hat V_{en}$),
electron-electron Coulomb interaction with the implicitly treated
electrons that gives rise to Coulomb ($\hat J$), and exchange
contribution ($\hat K$). The Coulomb contribution bears a resemblance
to the classical Coulomb interaction between an electron and the mean
electrostatic charge distribution of the implicitly electrons. In
contrast, exchange contribution is purely quantum. The form of the
Hamiltonian in Eq. \eqref{eq:hamiltonian_1e_mf} is analogous to the MF
one-electron Hamiltonian used in Hartree-Fock method.

\begin{table}[t!]
\caption{The directional renormalization factors for the hybridization
elements $t_{pd}$, the effective $(dd\sigma)$ hopping $t_\gamma$ which
depends on bond direction $\gamma$, the energies of final states for
charge transfer excitation $\Delta$, and for interionic Mn$-$Mn charge
excitations, $\varepsilon_n$ with $n=1,\dots,4$.}
\begin{ruledtabular}
\begin{tabular}{ccccc}
                & excited state        & & \multicolumn{2}{c}{bond renormalization} \\
   energy       &     Mn$-$O$-$Mn      & & $\langle ij\rangle{\parallel}a(b)$
                                         & $\langle ij\rangle{\parallel}c$  \\
\hline
  $t_{pd}$      &                      &          &  94.8\% & 111.5\% \\
  $t_\gamma$    &                      &          &  90.3\% & 123.7\% \\
                &  Mn--O  excitation   & $(3d)^5$ &         &         \\
  $\Delta$      & $(3d)^5(2p)^5(3d)^4$ & $^4A_1$  &  99.5\% & 100.5\% \\
                &  Mn--Mn excitation   &          &         &         \\
$\varepsilon_1$ & $(3d)^5(2p)^6(3d)^3$ & $^6A_1$  & 102.3\% &  95.6\% \\
$\varepsilon_2$ & $(3d)^5(2p)^6(3d)^3$ & $^4A_1$  & 101.0\% &  98.1\% \\
$\varepsilon_3$ & $(3d)^5(2p)^6(3d)^3$ & $^4E$    & 100.9\% &  98.3\% \\
$\varepsilon_4$ & $(3d)^5(2p)^6(3d)^3$ & $^4A_2$  & 100.7\% &  98.6\% \\
 \end{tabular}
\end{ruledtabular}
\label{tab:renormalization}
\end{table}

Having the concrete form of the orbitals and after identifying the
implicitly treated electrons, we were able to evaluate the value of
$t_{pd}$ according to Eq. (\ref{eq:t_pd_definition}). We constructed
the cluster of many unit cells to mimic the crystal environment. Then,
we invoked the approximation in which all implicitly treated electrons
localized at atoms different than the hopping spots are treated as
classical point charges (unit minus charges localized on the atomic
centers). The obtained results for the actual bond lengths in the $ab$
planes and along the $c$ axis are (in eV):
\begin{equation}
 t_{pd}(1.995\text{\AA{}}) = 2.00,
 \quad \quad
 t_{pd}(1.925\text{\AA{}}) = 2.35.
\label{tpd}
\end{equation}
These changes are in excellent agreement with the semi-empirical law of
Harrison which gives $t_{pd}\propto d_\gamma^{-4}$ for a Mn$-$O bond of
length $d_\gamma$ \cite{Har05}. For a shorter bond along the $c$ axis
of 97.72\%, one finds the increase of $t_{pd}$ to 109.7\%, while the
Harrison's law gives 111.4\%. For a bond longer by 101.27\% in the $ab$
plane one finds $t_{pd}$ decreased to 94.8\% while the Harrison's law
predicts 95.1\%. Indeed, these results verify the semi-empirical
Harrison's law and one finds a significant difference in the $t_{pd}$
values (\ref{tpd}). A kinetic energy contribution is roughly equal to
$5.0$ eV and an electrostatic potential contribution is roughly equal
to $-3.0$ eV (it includes the repulsive interaction with core electrons).
In addition to this we found weak ($-0.3$ eV) exchange contribution.
We compare these values with the value obtained for the cubic geometry
(2.11 eV) to get the $t_{pd}$ renormalization factors that should be
imposed when changing the cubic model into tetragonal one, the
first row in Table~\ref{tab:renormalization}.

Due to the large difference in the values of $t_{pd}$, the effective
hopping elements $t_{ab}$ and $t_c$ for the bonds ${\parallel}ab$ and
${\parallel}c$, respectively, exhibit remarkably large anisotropy and
$t_c/t_{ab}=1.37$, see Table \ref{tab:renormalization}. This large
difference may be obtained under neglect of the small anisotropy of CT
excitations, see the next Section, and modifies even more the
anisotropic $e_g$ exchange interactions due to their quadratic
dependence on the hopping, $J_{\gamma}^e\propto 4t_\gamma^2/U$,
and thus influences the spin-orbital order.

\subsection{Renormalization of the excitation energies}
\label{sec:exci}

In the superexchange approach all the virtual excited states are CT
states (in sense that there is a surplus positive charge located in
one lattice node accompanied by a surplus negative charge located in
another lattice node).
Energy of a CT configuration may be roughly divided into three parts:
the first part describes ionization energy of the first CT state
moiety,
the second part describes electron affinity of the second CT state
moiety, and
the third part describes an internal Coulomb interaction between the
surplus charges as well as their Coulomb interaction with
a surrounding charge distribution.
The first two parts are CT inter-moiety separation independent.
On the other hand, the third part is directly a crystal geometry
dependent part. For example, the Coulomb interaction between the
surplus charges is inversely proportional to inter-moiety distance
(from the simplest and approximate point of view).

Let us consider one of the virtual CT states: $(3d)^5(2p)^6(3d)^3$. In
this state the surplus unit negative charge is localized on the first
Mn atom and the surplus unit positive charge is localized on the second
Mn atom (in other words: it is the configuration Mn$^{2+}-$Mn$^{4+}$).
For Mn ions the $(3d)^5$ configuration may be realized by four
electronic terms:
$^6A_1$, $^4A_1$, $^4E$, and $^4A_2$, with the excitation energies
\cite{Kov10}: 1.93~eV, 4.52~eV, 4.86~eV and 6.24~eV,
using the parameters of Table \ref{tab:para}.
In the reference cubic geometry \mbox{Mn$-$Mn} distance is equal to
$3.94$~\AA{} and the Coulomb interaction energy between surplus charges
is equal to $-3.655$~eV. In the tetragonal geometry, in which in one
crystallographic direction \mbox{Mn$-$Mn} distances are equal to
$3.99$~\AA{} or $3.85$~\AA{}, and the corresponding Coulomb interaction
energy is $-3.609$ eV or $-3.740$ eV, respectively.

In addition to this the surplus charges --- surrounding ion
interactions should be considered; however in this case they cancel out
as the surplus charges are located on the equivalent lattice positions.
It means that the contribution to total excitation energy is bigger for
$0.045$~eV (or less for $-0.085$~eV). For example, for $^6A_1$
excitation this change gives rise to the renormalization of the total
excitation energy by the factor 102.4\% (or 95.6\%). The renormalization
results are collected in Table \ref{tab:renormalization}.

Finally, there is as well the $(3d)^5(2p)^5(3d)^4$ configuration
standing for a CT excitation. Its overall excitation energy is equal to
5.5 eV \cite{Boc92}. The corresponding electrostatic calculation is more
involved in this case: as the opposite surplus charges are not located
at equivalent lattice positions, one has to add surplus charges ---
surrounding ion interactions energy. The surplus positive charge
(located on an oxygen anion) has bigger energy if the hopping goes in
the shortened direction. This effect is almost exactly compensated by
the energy gain due two surplus charges stabilization (that appears if
they are closer). The net result is that the excitation energies are
barely split and corresponding excitation energies renormalization
factor is close to unity.

\section{The tetragonal model at $T=0$}
\label{sec:pre}

\subsection{Orbital mixing angle and exchange constants}
\label{sec:T=0}

The predictions of the tetragonal model Eq. (\ref{som}) were
investigated using on-site MF approximation and compared to the results
obtained for the cubic model. We decided to implement the extensions
accounting for tetragonal CF together with the directional
renormalization of hopping integral and neglect the weaker
electrostatic effects. We focus in this Section on the results obtained
at $T=0$. When temperature increases, the tetragonal deformation varies
with $T$ and a purely electronic model cannot be applied,
see the Appendix.

The model parameter values were fitted to reproduce the available
experimental observations. We used effective hopping integral
$t = t_{pd}^2 / \Delta = 0.5$~eV (for CT energy $\Delta=5.5$ eV
\cite{Boc92} this corresponds to $t_{pd}=1.66$~eV, reasonably close
to the value 1.5~eV used before \cite{Fei99}). The
temperature of the structural transition $T_{\rm OO}$ is only a
fraction of the orbital exchange interaction \cite{Cza17} and alone
would not explain the high value of $T_{\rm OO}$
\cite{Fei98,Fei99,Oka02,Pav10}. Thus we included the JT effective
parameter $\kappa$ which was chosen 4~meV (this value is lower than
that used before \cite{Snm16} as a consequence of the increased values
of the hopping parameters $t_\gamma$ and enhanced superexchange).
At the end we used $E_z=300$~meV, in line with our external
quantum chemical calculations (\ref{Ez}), see Sec. \ref{sec:tetra}.
Other physical parameters were the same as deduced for the cubic model,
see Table~\ref{tab:para}.

The obtained results are remarkably satisfactory --- they reproduce all
experimental data with reasonable accuracy. Obtained the orbital mixing
angle $\vartheta$ (\ref{mixing}) is equal to 106$^\degree$ and stays
within excellent agreement with the experimental value 108$^\degree$.
We recall that the orbital mixing angle was notoriously underestimated
in the cubic model (we reported before 84$^\degree$ \cite{Snm16}). This
improvement is mainly due to the tetragonal $e_g$ orbital splitting.
The estimated value $E_z=300$~meV (\ref{Ez}) fits perfectly.

Also the predicted values of the spin exchange constants,
$J_{ab}=-1.7$~meV and $J_c=+0.8$~meV (both at $T=0$), are reasonably
close to the experimental values: $-1.7$~meV and $1.2$~meV \cite{Mou96}.
The results depend strongly on the introduced renormalization of
$t_{pd}$ hopping integral (\ref{tpd}) that gives rise to immense
$J_{ab}$ and $J_c$ renormalization factors (-29\% and 51\%,
respectively). Without them the $|J_{ab}|/J_c$ ratio would be incorrect.
Although the values fit reasonably well, they are numerically uncertain
as they are an algebraic sum of a few big (up to 9~meV) contributions
of opposite sign.
(The spin exchange constants consist of a strong FM superexchange term
for the lowest energy $^6A_1$-excitation and several smaller AF terms.)

\subsection{Discussion}
\label{sec:dis}

As one may expect that the tetragonal crystal field $E_z$ (\ref{Hz}) is
the most important correction to introduce, we tentatively investigated
it at first as if it was the only needed correction (and neglected the
other two proposed later). Within this approach, the discussed model
was essentially the same as in Ref. \cite{Fei99}. At this stage we
adopted, as a reference starting point, the cubic model parameter
values from our earlier work \cite{Snm16} that was devoted to
(implicitly) cubic model. The phase diagram was examined against the
$e_g$ orbital splitting, i.e., tetragonal crystal field $E_z$, see
Fig.~\ref{fig:EzInpact}. We have found that too strong crystal field
(greater than 0.2 eV) could have destabilized the observed $A$-AF
state. This transition bears great resemblance to the transition
between $A$-AF phase and $G$-AF$x$ phase described in
Ref.~\cite{Fei99}. The only difference is that the $G$-AF phase with
adjusted orbitals arises in addition between the two phases
(for very tiny range of $E_z$).

Then, the dependence between orbital mixing angle and the tetragonal
crystal field magnitude was plotted. As expected, tetragonal crystal
field enhances the amplitude of the $|x^2-y^2\rangle$ orbital state
(the $|3z^2-r^2\rangle$ counterpart contributes mainly to excited
states at both sublattices). We found that the orbital mixing angle
deduced experimentally for the orbital ordered state is reproduced when
$E_z\simeq 200$~meV. At the end of this part we carried out independent
quantum chemical calculations to evaluate the value of $E_z$. These
calculations suggest $E_z=300$~meV (\ref{Ez}), and we conclude that the
entire picture is consistent (and more realistic than the one offered
by the implicitly cubic model).
Thus we suggest that it may be used as a starting point to include
finer corrections (considering directional renormalization).

Then, to establish an even more reliable Hamiltonian we investigated
the origin of superexchange terms. Improvements include the directional
dependence of the hopping integral $t_{pd}$ values as well as the
excitation energies of virtual charge excited states. The directional
deviations of the values introduce directional renormalization factors
of Hamiltonian terms that extend the cubic Hamiltonian analyzed earlier.
We employed chemical models to evaluate the renormalization factors in
an approximate way using first principle methods. The raw, classical
calculations suggest that the changes of excitations energies are
small --- only up to a few percent --- and may be neglected on the
presented qualitative level of theory.

On the other hand, the changes of hopping integrals (\ref{tpd}) are
considerable. We found the following $t_{pd}$ values: 2.11~eV for the
cubic geometry and $t_{pd}\simeq 2.0$~eV or $t_{pd}\simeq 2.35$~eV for
tetragonal geometry (depending on the bond direction), see Eqs.
(\ref{tpd}). The anisotropy of $t_{pd}$ is enhanced by a factor of two
in the Mn$-$Mn effective hopping, see Eq. (\ref{tpd2}). Note that the
values of $t_{pd}$ are somewhat larger than those used earlier (1.5~eV
in Ref. \cite{Fei99}). The increase of $t_{pd}$ may be rationalized by
the fact that Mn$-$O bonds in our conceived tetragonal model are shorter
than in the real crystal (due to the Mn$-$O$-$Mn bridges bends as
described in the first paragraph of this Section). For the real Mn$-$O
bond lengths, obtained $t_{pd}$ values would be smaller.

\section{Summary}
\label{sec:summa}

In this work we analyzed diverse physical aspects that emerge due to
the non-equivalence of all three crystallographic directions in the
observed perovskite structure of LaMnO$_3$ crystal. In order to make
the analysis conclusive the conceived tetragonal LaMnO$_3$ crystal
model was introduced and investigated. We demonstrated that it offers
a better overall description of the ground state in comparison with
the models used earlier (that implicitly assume that LaMnO$_3$ crystal
has a cubic perovskite structure). Despite apparent improvements, the
introduced model adopts complicated structural effects only partially.
In fact, the real LaMnO$_3$ crystal is non-tetragonal due to additional
MnO$_6$-octahedra tilts. The bridging oxygen atoms do not lie precisely
on the line segments spanning the bridged manganese ion pairs --- the
oxygen atoms are translated aside the segments. The Mn$-$O$-$Mn bridges
bend so that the Mn$-$O bonds are less strained. We believe that the
corrections to the crystal electronic structure that stem from the
tilts are of less importance and their qualitative description goes
beyond the scope of this work.

Further support for the tetragonal model introduced here follows from
its comparison with experiment, both at $T=0$ and at finite temperature.
The predicted electronic ground state was investigated in terms of the
observables that may be compared with experiment: the spin exchange
constants and orbital mixing angle. We found excellent agreement with
the experimental data when the parameters of Table I are used, but we
had to modify two of them: $t=0.5$~eV and $\kappa=4$~meV.
This agreement appears even surprising as both the orbital angle and
the exchange constants are well reproduced, in contrast to earlier
studies \cite{Ole05,Snm16}. Indeed, such a satisfactory result is very
hard to obtain without the described directional renormalizations of
$t_{pd}$ which directly influence the $J_c/J_{ab}$ ratio. At finite
temperature we established tentative link between structural
deformation and the orbital order parameter. Using the tetragonal model
Hamiltonian we obtained the predicted temperatures of spin (magnetic)
and orbital phase transitions, see the Appendix. We observe no changes
of transition temperatures due to the directional corrections, however
the critical exponent for orbital phase transition is modified as
the tetragonal distortion is reduced together with orbital order.

Summarizing, we established the tetragonal model of LaMnO$_3$ by
evaluating the corrections necessary to improve the model (\ref{som})
derived initially for the perovskite structure. We conclude that for
LaMnO$_3$ at least two physical effects connected with the inequality
of Mn$-$O$-$Mn bridge distances are important and have to be included
when carrying out quantitative calculations:
(i)~the tetragonal crystal field $E_z$ (\ref{Ez}) and
(ii) direction-dependent renormalization of $p-d$ hybridization
$t_{pd}$ (\ref{tpd}).
The first one is necessary to obtain the experimentally observed form
of the occupied $e_g$ orbitals in the ground state, i.e., the correct
value of the orbital mixing angle (\ref{mixing}) for the occupied
states. The second one is responsible for the anisotropy of the
dominating $e_g$ part of superexchange and is crucial to reproduce
the observed ratio of spin exchange constants, $J_c/J_{ab}$. These
effects are far stronger than the systematic corrections beyond the
simplest on-site mean field described in our previous work
\cite{Snm16}. We suggest that similar modifications of the ground state
may be found in other correlated insulators when the present procedure
was repeated for realistic crystal structure of similar transition
metal compounds, for instance for KCuF$_3$
(although some technical problems could occur).

\acknowledgments
We thank K. Ro\'sciszewski for insightful discussions.
We kindly acknowledge support by Narodowe Centrum Nauki
(NCN, National Science Centre, Poland)
under Project No.~2016/23/B/ST3/00839.

\appendix*
\section{Tetragonal model at finite temperature}
\label{sec:state}

The influence of the introduced parameter renormalization on (spin and
orbital) transition temperatures is of particular interest. To analyze
transition temperatures, on-site MF calculations were performed at
finite temperature. We emphasize that as temperature increases the
tetragonal deformation decreases, and starting from the orbital
transition temperature $T_{\rm OO}$ the the orbital order disappears
\cite{Mur98} and cubic model is valid again.
Therefore the thermal dependence of deformation magnitude has to be
included when performing on-site MF calculations at finite temperature.

The proposed model (\ref{som}) is purely electronic and the structural
deformation magnitude is one of its external parameters. The values of
all the parameters, as well as their temperature dependence, have to be
introduced as input. However, it is misleading to simply use a fixed
function of temperature that describes the deformation, as the
electronic state and crystal geometry are strictly interrelated.
For instance, the calculations have to respect the physical requirement
that the orbital order phase transition and structural phase transition
take place simultaneously. To assure the consistency
(between electronic and structural degrees of freedom) we use a
strategy of dynamical updating of deformation magnitude as a function
of electronic state. Although there is no direct way to gain
information about the unit cell parameters form electronic state we
propose to do it straightforwardly. We postulate that the deformation
is proportional to the on-site orbital order parameter $r(T)$.
Hence we arrive at the heuristic equations:
\begin{align}
E_z(T)&= r(T) E_z(0),
 \\
t_{pd,\gamma}(d_\gamma,T) &= t_{pd}^\square+r(T)\left[t_{pd,\gamma}(d_\gamma,0)-t_{pd}^\square\right].
\end{align}
where $t_{pd}^\square$ denotes the reference value for a cubic crystal.
Our strategy may be verified \textit{a posteriori} by comparison of the
obtained functional dependence of crystal deformation with available
crystallographic data.

The tetragonal model gives in mean field approximation the following
transition temperatures: 160~K for the magnetic transition and 1615K
for the simultaneous orbital/structural transition.
As the experimental values are equal to 140K and 750~K the theory
overestimates them by 115\% and 215\%. Indeed, one may expect
enormous overestimation due to crude on-site MF approximation,
and the overestimation factors up to 200\% for the
orbital pseudospin $\tau=1/2$ are acceptable.

\begin{figure}[t!]
 \begin{center}
 \includegraphics[width=\columnwidth]{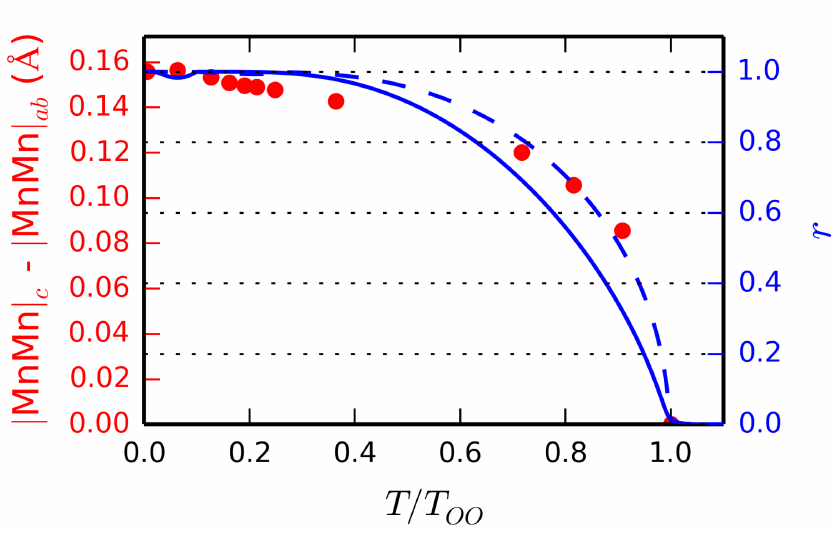}
 \end{center}
 \caption{
Comparison of thermal dependences of calculated orbital order parameter
$r(T)$ (blue solid curve for tetragonal model and blue dashed curve for
cubic model) with experimental data of tetragonal deformation magnitude
(red points). The experimental data are from \cite{Crv98}.}
 \label{fig:r_and_deformation}
\end{figure}

Finally, the heuristic assumption about the form of the interrelation
between the electronic state and the structural deformation magnitude
should be checked. We compare the shape of the temperature dependence
of measured deformation magnitude and the orbital order parameter
$r(T)$ extracted from the theory. In order to compare the shape, both
functions are expressed in terms of $T/T_{\rm OO}$, where $T_{\rm OO}$
is the corresponding orbital transition temperature (obtained within
the present theory), see Fig.~\ref{fig:r_and_deformation}. Both curves
are qualitatively similar and the agreement is fair. The curve 
obtained for the tetragonal model fits better than the one for the 
cubic model in low-to-medium temperature sector as it bends much
stronger for $0.3<T/T_{\rm OO}<0.7$. However the curve for cubic model
seems to offer a better approximation for the critical exponent $\beta$
as long as the electronic model is considered. The qualitative 
correction to MF value seems to be a consequence of the temperature
dependence of tetragonal distortion which modifies the orbital 
transition. We also note that remarkable fast decrease of the 
tetragonal deformation for $T/T_{\rm OO}<0.4$ cannot be reproduced in 
the electronic model --- it may be expected that acoustic phonons are 
of importance in this regime and the deformation decreases faster than 
predicted in mean field theory. 
This question remains open for future studies.

\begin{thebibliography}{99}


\bibitem{Ima98} M. Imada, A. Fujimori, and Y. Tokura,
                   Metal-insulator transitions,
                   Rev. Mod. Phys. \textbf{70}, 1039 (1998).

\bibitem{Kug82} K. I. Kugel and D. I. Khomskii,
                   The Jahn-Teller effect and magnetism:
                   Transition metal compounds,
                   Usp. Fiz. Nauk \textbf{136}, 621 (1982)
                  [Sov. Phys. Usp. \textbf{25}, 231 (1982)].

\bibitem{Mil96} A. J. Millis,
                   Cooperative Jahn-Teller effect and electron-phonon
                   coupling in La$_{1-x}$A$_x$MnO$_3$,
                   Phys. Rev. B \textbf{53}, 8434 (1996).

\bibitem{Vol10} I. Leonov, D. Korotin, N. Binggeli, V. I. Anisimov, and D. Vollhardt,
                   Computation of correlation-induced atomic
                   displacements and structural transformations
                   in paramagnetic KCuF$_3$ and LaMnO$_3$,
                   Phys. Rev. B \textbf{81}, 075109 (2010).

\bibitem{Kha05} G. Khaliullin,
                   Prog. Theor. Phys. Suppl. \textbf{160}, 155 (2005).

\bibitem{Ole05} A. M. Ole\'s, G. Khaliullin, P. Horsch, and L. F. Feiner,
                   Fingerprints of spin-orbital physics in cubic Mott insulators:
                   Magnetic exchange interactions and optical spectral weights,
                   Phys. Rev. B \textbf{72}, 214431 (2005).

\bibitem{Rei05} A. Reitsma, L. F. Feiner, and A. M. Ole\'s,
                   Orbital and spin physics in LiNiO$_2$ and NaNiO$_2$,
                   New J. Phys. \textbf{7}, 121 (2005).

\bibitem{Mila}  A. Smerald and F. Mila,
                   Disorder-Driven Spin-Orbital Liquid Behavior
                   in the Ba$_3$XSb$_2$O$_9$ Materials,
                   Phys. Rev. Lett. \textbf{115}, 147202 (2015);
                   Exploring the spin-orbital ground state of Ba$_3$XSb$_2$O$_9$,
                   Phys. Rev. B \textbf{90}, 094422 (2014).

\bibitem{Karlo} P. Corboz, M. Lajk\'o, A. M. La\"uchli, K. Penc, and F.~Mila,
                   Spin-Orbital Quantum Liquid on the Honeycomb Lattice,
                   Phys. Rev. X \textbf{2}, 041013 (2012).

\bibitem{Brz15} W. Brzezicki, A. M. Ole\'s, and M. Cuoco,
                   Spin-orbital order modified by orbital dilution
                   in transition-metal oxides: From spin defects
                   to frustrated spins polarizing host orbitals,
                   Phys. Rev. X \textbf{5}, 011037 (2015).

\bibitem{Brz16} W. Brzezicki, M. Cuoco, and A. M. Ole\'s,
                   Novel Spin-Orbital Phases Induced by Orbital Dilution,
                   J. Supercond. Nov. Magn. \textbf{29}, 563 (2016);
                   Exotic Spin-Orbital Physics in Hybrid Oxides,
                   \textit{ibid.} \textbf{30}, 129 (2017).

\bibitem{Dag01} E. Dagotto, T. Hotta, and A. Moreo,
                   Colossal Magnetoresistant Materials:
                   The Key Role of Phase Separation,
                   Phys. Rep. \textbf{344}, 1 (2001);
                E. Dagotto,
                   Open Questions in CMR Manganites, Relevance of Clustered
                   States and Analogies with Other Compounds Including the Cuprates,
                   New J. Phys. \textbf{7}, 67 (2005).

\bibitem{Tok06} Y. Tokura,
                   Critical Features of Colossal Magnetoresistive Manganites,
                   Rep. Prog. Phys. \textbf{69}, 797 (2006).

\bibitem{Fei99} L. F. Feiner and A. M. Ole\'s,
                   Electronic Origin of Magnetic and Orbital
                   Ordering in Insulating LaMnO$_3$,
                   Phys. Rev. B \textbf{59},   3295 (1999).

\bibitem{Kov10} N. N. Kovaleva, A. M. Ole\'s, A. M. Balbashov, A. Maljuk,
                   D.~N.~Argyriou, G. Khaliullin, and B. Keimer,
                   Low-energy Mott-Hubbard excitations in LaMnO$_3$ probed
                   by optical ellipsometry,
                   Phys. Rev. B \textbf{81}, 235130 (2010).

\bibitem{Snm16} M. Snamina and A. M. Ole\'s,
                   Spin-orbital order in the undoped manganite LaMnO$_3$
                   at finite temperature,
                   Phys. Rev. B \textbf{94}, 214426 (2016).

\bibitem{Ole12} A. M. Ole\'s,
                   Fingerprints of spin-orbital entanglement
                   in transition metal oxides,
                   J. Phys.: Condens. Matter \textbf{24}, 313201 (2012);
                   Frustration and entanglement
                   in compass and spin-orbital models,
                   Acta~Phys. Polon. A \textbf{127}, 163 (2015).

\bibitem{You15} W. Brzezicki, J. Dziarmaga, and A. M. Ole\'s,
                   Noncollinear magnetic order stabilized by entangled
                   spin-orbital fluctuations,
                   Phys. Rev. Lett. \textbf{109}, 237201 (2012);
                   Exotic Spin Orders driven by orbital fluctuations
                   in the Kugel-Khomskii Model,
                   Phys. Rev. B \textbf{87}, 064407 (2013);
                   Topological Order in an Entangled Spin-Orbital
                   SU(2)$\otimes XY$ Ring,
                   Phys. Rev. Lett. \textbf{112}, 117204 (2014).

\bibitem{Fei98} D. Feinberg, P. Germain, M. Grilli, and G. Seibold,
                   Joint superexchange-Jahn-Teller mechanism
                   for layered antiferromagnetism in LaMnO$_3$,
                   Phys. Rev. B \textbf{57}, R5583 (1998).

\bibitem{Oka02} S. Okamoto, S. Ishihara, and S. Maekawa,
                   Orbital ordering in LaMnO$_3$:
                   Electron-electron and electron-lattice interactions,
                   Phys. Rev. B  \textbf{65}, 144403 (2002).

\bibitem{Pav10} Eva Pavarini and Erik Koch,
                   Origin of Jahn-Teller Distortion
                   and Orbital Order in LaMnO$_3$,
                   Phys. Rev. Lett. \textbf{104}, 086402 (2010).

\bibitem{Fle12} A. Flesch, G. Zhang, E. Koch, and E. Pavarini,
                   Orbital-order melting in rare-earth manganites:
                   Role of superexchange,
                   Phys. Rev. B \textbf{85}, 035124 (2012).

\bibitem{Saw97} H. Sawada, Y. Morikawa, K. Terakura, and N.~Hamada,
                   Jahn-Teller distortion and magnetic structures in LaMnO$_3$,
                   Phys. Rev. B \textbf{57}, R3189 (1998).

\bibitem{Crv98} J. Rodr\'iguez-Carvajal, M. Hennion, F. Moussa,
                   A. H. Moudden, L. Pinsard, and A. Revcolevschi,
                   Neutron-diffraction study of the Jahn-Teller transition
                   in stoichiometric LaMnO$_3$,
                   Phys. Rev. B \textbf{57}, R3189 (1998).

\bibitem{Miz99} T. Mizokawa, D. I. Khomskii, and G. A. Sawatzky,
                   Interplay between orbital ordering and lattice distortions
                   in YVO$_3$, YTiO$_3$, and LaMnO$_3$,
                   Phys. Rev. B \textbf{60}, 7309 (1999).

\bibitem{Alo00} J. A. Alonso, M. J. Martinez-Lope, M. T. Casais,
                   and M.~T.~Fernandez-Diaz,
                   Evolution of the Jahn-Teller distortion of MnO$_6$
                   octahedra in $R$MnO$_3$ perovskites
                   ($R$=Pr, Nd, Dy, Tb, Ho, Er, Y): A neutron diffraction study,
                   Inorganic Chem. \textbf{39}, 917 (2000).

\bibitem{Kho14} D. I. Khomskii, \textit{Transition Metal Compounds}
                   (Cambridge University Press, Cambridge, 2014).

\bibitem{Dag04} M. Daghofer, A. M. Ole\'s, and W. von der Linden,
                   Orbital polarons versus itinerant $e_g$ electrons
                   in doped manganites,
                   Phys. Rev. B \textbf{70}, 184430 (2004).

\bibitem{Hua97} Q. Huang, A. Santoro, J. W. Lynn, R. W. Erwin, J. A. Borchers,
                   J. L. Peng, and R. L. Greene,
                   Structure and magnetic order in undoped lanthanum manganite,
                   Phys. Rev. B \textbf{55}, 14987 (1997).

\bibitem{Bec93} A. D. Becke,
                   Density-functional thermochemistry:
                   III.~The role of exact exchange,
                   J. Chem. Phys. \textbf{98}, 5648 (1993).

\bibitem{Ste94} P. J. Stephens, F. J. Devlin, C. F. Chabalowski, and M.~J.~Frisch,
                   Ab Initio Calculation of Vibrational Absorption and Circular
                   Dichroism Spectra Using Density Functional Force Fields,
                   J. Chem. Phys. \textbf{98}, 11623 (1994).

\bibitem{Zaa88} J. Zaanen and A. M. Ole\'s,
                   Canonical Perturbation Theory and the Two-Band Model
                   for High-$T_c$ Superconductors,
                   Phys. Rev. B \textbf{37}, 9423 (1988).

\bibitem{Har05} W. A. Harrison, \textit{Elementary Electronic Structure}
                   (World Scientific, Singapore, 2005).

\bibitem{Boc92} A. E. Bocquet, T. Mizokawa, T. Saitoh, H. Namatame, and A. Fujimori,
                   Electronic structure of $3d$-transition-metal compounds
                   by analysis of the $2p$ core-level photoemission spectra,
                   Phys. Rev. B \textbf{46}, 3771 (1992).

\bibitem{Cza17} P. Czarnik, J. Dziarmaga, and A. M. Ole\'s,
                   Overcoming the Sign Problem at Finite Temperature:
                   Quantum Tensor Network for the Orbital $e_g$ Model
                   on an Infinite Square Lattice,
                   Phys. Rev. B \textbf{96}, 014420 (2017).

\bibitem{Mou96} F. Moussa, M. Hennion, J. Rodr\'iguez-Carvajal,
                   H.~Moudden, L. Pinsard, and A. Revcolevschi,
                   Spin waves in the antiferromagnet perovskite
                   LaMnO$_3$: A~neutron-scattering study,
                   Phys. Rev. B \textbf{54}, 15149 (1996);
                G. Biotteau, M. Hennion, F. Moussa,
                   J.~Rodr\'iguez-Carvajal, L. Pinsard,
                   A. Revcolevschi, Y.~M.~Mukovskii, and D. Shulyatev,
                   Approach to the metal-insulator transition in
                   La$_{1-x}$Ca$_x$MnO$_3$ ($0<\sim x<\sim 0.2$):
                   Magnetic inhomogeneity and spin-wave anomaly,
                   \textit{ibid.} \textbf{64}, 104421 (2001).

\bibitem{Mur98} Y. Murakami, J. P. Hill, D. Gibbs, M. Blume, I. Koyama,
                   M. Tanaka, H. Kawata, T. Arima, Y. Tokura, K. Hirota,
                   and Y. Endoh,
                   Resonant X-Ray Scattering from Orbital Ordering in LaMnO$_3$,
                   Phys. Rev. Lett. \textbf{81}, 582 (1998).



\end{thebibliography}
\end{document}